\documentclass[aps,reprint,prd,showpacs,amsmath,amssymb,nofootinbib]{revtex4-1} 
\usepackage{graphicx}
\usepackage{color}
\usepackage{times}
\usepackage{inputenc}
\usepackage{bm}
\usepackage{multirow}
\usepackage{float}
\usepackage{url}
\usepackage{natbib}
\usepackage{tensor}
\usepackage{bbm}
\usepackage[colorlinks=true,citecolor=blue,urlcolor=blue,linkcolor=red]{hyperref}
\usepackage[normalem]{ulem}
\usepackage{verbatim}

\def\fm3{\;\text{fm}^{-3}}
\def\mev{\;\text{MeV}}

\begin{document}
\title{Diquark and chiral condensates in a self-consistent NJL-type model}
\author{Wen-Li Yuan$^{1,2}$}
\email{wlyuan@smail.nju.edu.cn}
\author{Jingyi Chao$^{3}$}
\email{chaojingyi@jxnu.edu.cn}
\author{Ang Li$^{2}$}
\email{liang@xmu.edu.cn}

\affiliation{$^1$Department of Physics, Nanjing University, Nanjing 210093, China;\\ 
$^2$Department of Astronomy, Xiamen University, Xiamen, Fujian 361005, China;\\ 
$^3$College of Physics and Communication Electronics, Jiangxi Normal University, Nanchang 330022, China
}

\begin{abstract}
In this work, a modified two-flavor NJL-type model is utilized, in which the contact current-current interaction is Fierz-transformed into quark-antiquark interactions and quark-quark interactions, which are directly related to the chiral condensate and diquark condensate, respectively. Under mean-field approximation, the chiral condensate and the diquark condensate are studied on the same footing. We discuss in detail the competition between the chiral condensate and the diquark condensate, which are exclusively paired with u and d quarks, while also investigating the order of the chiral phase transition through an analysis of the resulting chiral susceptibility.
\end{abstract}


\maketitle

\section{Introduction}\label{:intro} 
Quarks may readily pair up~\cite{1975PhRvL..34.1353C,1977NuPhB.129..390B,1984PhR...107..325B} in high-density quark matter, forming Cooper pairs, similar to electrons in metals~\cite{1957PhRv..106..162B,1957PhRv..108.1175B}. 
Because quantum chromodynamics (QCD), the first-principle theory of quarks, is an asymptotically free theory~\cite{1973PhRvL..30.1343G,1973PhRvL..30.1346P}, color superconducting (CSC) quark matter at asymptotically high densities can be analyzed through a well-controlled weak coupling approach~\cite{1995PhLB..350..163I,1999PhRvD..60i4013P,1999PhRvD..60k4033S,2000PhRvD..61g4017P}. Though the superconducting gap and critical temperature calculated using this approach are rather small, around $1 \mev$~\cite{1984PhR...107..325B}, the corresponding pairing gaps of strongly interacting matter, calculated from e.g., the Nambu-Jona–Lasinio (NJL) model, were found to be far greater, on the order of $100\mev$~\cite{1998PhLB..422..247A,1998PhRvL..81...53R,1999PhRvC..60e5205S}. 
These studies highlighted the relevance of CSC matter for the QCD phase diagram, and recognized its potential significance to neutron star studies~\cite{2002JHEP...06..031A,2003PhRvD..67j3004S,2003PhLB..562..153B,2004PhLB..595...36B,2015PhRvD..91d5003K,2019ApJ...885...42B,2021ApJ...917L..22M} and possibly even in heavy-ion collisions~\cite{1995PhLB..350..163I,1999PhRvD..60i4013P,1999PhRvD..59i4019S}. The possible existence of CSC quark matter thus became an appealing field of research~\cite{1999NuPhB.537..443A,1999PhRvD..59i4019S,2000NuPhS..83..103S,2002PhRvD..65g6012H,2003PhLB..564..205S,2020PhRvD.102e4028S,2021ApJ...917L..22M}. Not only does CSC matter enrich the QCD phase structure~\cite{1999NuPhB.538..215B,2003PhLB..564..205S,2005PhRvD..72c4004R,2005PhRvD..72f5020B}, its presence can also contribute to a deeper understanding of heavy-ion collisions and neutron stars. For a comprehensive review of this topic, see Refs.~\cite{2001afpp.book.2061R,2004PrPNP..52..197R,2008RvMP...80.1455A,2014RvMP...86..509A}.

For the methods to calculate the magnitude of the gap parameters in the CSC phases~\cite{2001afpp.book.2061R}, there are two distinct paths that have been followed. One path commences from first principles and relies on the property of asymptotic freedom of QCD, by utilizing renormalization group techniques, or through the Schwinger-Dyson equation~\cite{1999PhRvD..59i4019S,1999PhLB..449..281E,2000PhRvD..61e6001H}. The other path involves a semiphenomenological method, such as the above-mentioned NJL model~\cite{1992RvMP...64..649K,2005PhR...407..205B}, in which the interaction between quarks is substituted by a four-fermion interaction, which originates from instanton exchange~\cite{1998PhLB..422..247A,1998PhRvL..81...53R} or induced by single-gluon exchange~\cite{1999NuPhB.537..443A,1999NuPhB.558..219A}. 

Quarks differ from electrons in that they carry color and flavor degrees of freedom, which allows for numerous potential pairing patterns~\cite{1999NuPhB.538..215B,2003PhLB..564..205S,2005PhRvD..72c4004R,2005PhRvD..72f5020B}. 
At sufficiently high density and low temperature, the thermodynamically favored phase is the color-flavor-locked (CFL) phase~\cite{1999NuPhB.558..219A,1999NuPhB.537..443A,2019PhRvD..99c6004A}. This phase involves the pairing of lighter up (u) and down (d) quarks with heavier strange (s) quarks. 
And the gapless CFL (gCFL) phase~\cite{2004PhRvL..92v2001A} was proposed at some critical value of the strange quark mass.
 Additionally, other phases have been discovered at low temperatures, including the two-flavor color-superconducting (2SC) phase~\cite{1998PhLB..422..247A,2000PhRvD..61e6001H}, the gapless 2SC (g2SC) phase~\cite{2003PhLB..564..205S,2003NuPhA.729..835H}, a metallic CFL (mCFL) phase, and a uSC phase~\cite{2004PhRvL..93m2001I}. 
This work focuses on the 2SC phase in quark matter consisting of only u and d quarks. In this particular phase, the Fermi momenta of different quark flavors are approximately equal, and the u and d quarks form Cooper pairs in the color-antitriplet, flavor-singlet, spin-zero channel~\cite{1998PhLB..422..247A,1999PhRvD..59i4019S,2000PhRvD..61e1501P}. We will use NJL-type models to study the 2SC phase for potential applications in strongly interacting matter within compact stars.
 
In NJL-type models, for the contact current-current interaction, Fierz transformation can explicitly display all forms of quark-antiquark interaction channels $(\bar{q} q)$ and quark-quark interaction channels {$(q q)$}
~\cite{1992RvMP...64..649K,2005PhR...407..205B,2019ChPhC..43h4102W,2019PhRvD.100i4012Y,2020PhRvD.102e4028S}, which enables us to derive a general four-fermion interaction.
Under mean-field approximation, the quark-antiquark interaction channels produce the chiral condensate $\langle \bar{q} q \rangle$, which serves as the order parameter for the chiral phase transition. Meanwhile, the quark-quark interaction channels contribute to the diquark condensate $\langle qq \rangle$, which can be viewed as the order parameter for color superconductivity~\cite{2005PhR...407..205B,2005PhRvD..72c4004R,2005PhLB..611..137G}. As a result, the dynamical mass generation and diquark condensate are described self-consistently at equal levels. However, the weighting factor $\alpha$ cannot be well-defined under mean-field approximations and is typically adjusted within a range of $0$ to $1$ to be consistent with finite-density constraints.

Assuming equal contributions from the quark-antiquark interaction channels and Fierz-transformed diquark interaction channels, a coupling ratio of $h_1/g_s^{(0)}=3/4$ has been commonly adopted in previous studies~\cite{2005PhR...407..205B,2005PhRvD..72e4024S,2014RvMP...86..509A,2015PhRvD..92j5030C}, where $h_1$ and $g_s^{(0)}$ denote the coupling constants of the scalar diquark interaction and scalar quark-quark interaction, respectively. However, there is no physical basis for assuming this equal combination. Thus, we are also interested in utilizing the current modified NJL-type model to systematically investigate the effect of changing the $h_1/g_s^{(0)}$ ratio.
Additionally, at finite densities, the existence of Lorentz non-invariant expectation values becomes possible, and the ground state of 2SC phase is characterized by more condensates than merely the Lorentz-invariant ones $\Delta_1$ (the scalar diquark condensate)~\cite{2020PhRvD.102e4028S}. Therefore, we further take into account the vector diquark condensate $\Delta_2$, which is Lorentz non-invariant and transforms like the time component of a four-vector. Furthermore, vector interactions among quarks significantly impact the quark chemical potential and pressure, leading to substantial changes in the properties of the phase transition and the stiffness of equation of state~\cite{1996NuPhA.611..393B,2012PhRvD..85k4017S,2022PhRvD.105l3004Y}. Thus, we incorporate both scalar ($\bar{q}\lambda_a q$) and vector color-octet ($\bar{q}\gamma^{\mu} \lambda_a q$) channels in our present study. As demonstrated below, both scalar and vector color-octet interactions contribute to the differences in both the quark condensate and the quark number densities between paired and unpaired quarks. 

This paper is organized as follows. 
In Sec.~\ref{:model}, we introduce the effective Lagrangian of the NJL-type model. We derive both the chiral gap equation and the color superconducting gap equation in the mean-field approximation. In Sec.~\ref{:property}, we derive the thermodynamic potential and provide expressions for significant expectation values. In Sec.~\ref{:result}, we analyze the chiral and CSC phase transitions of high-density and zero-temperature QCD matter for various parameter values of $\alpha$. We summarize the results in Sec.~\ref{:summary}.

\section{The modified NJL-type Model}
\label{:model}
Assuming that gluon degrees of freedom can be frozen into point-like effective interactions between quarks, one can obtain the NJL-type models in which the interaction is replaced by a contact current-current interaction~\cite{1992RvMP...64..649K,1999PhRvC..60e5205S,2005PhR...407..205B} as follows:
\begin{equation}
\mathcal{L}=\bar{q}\left(i \gamma^{\mu} \partial_{\mu}-m + \mu \gamma_{0} \right) q+\mathcal{L}_{\text {int}}^{(4)}\ ,\label{eq1}
\end{equation}
where the interaction term between two currents is given by:
\begin{equation}
\mathcal{L}_{\text {int}}^{(4)}=-g\sum_{a=1}^{8}\left(\bar{q} \gamma^{\mu} \lambda_{a} q\right)^{2}.\label{eq2}
\end{equation}
Here, $\gamma^{\mu}$ are the Dirac gamma matrices and $\lambda_{a}$ are the generators of $\mathrm{SU(3)_c}$. $q$ is the quark field operator with color, flavor, and Dirac indices. $m$ is the diagonal mass matrix for quarks, which includes the bare quark masses and introduces a small explicit symmetry breaking.
In the following, we start with how to obtain the present version of the extended two-flavor NJL-type model and analyze which interactions we considered.

By applying the Fierz transformation to the two-quark-current interaction term in Eq.~(\ref{eq1}), 
it can be decomposed into $\bar{q}q$ and $qq$ channels respectively~\cite{1992RvMP...64..649K,2005PhR...407..205B}.
This allows us to combine the interactions with the Fierz-transformed ones using a weighting factor $\alpha$, thus enabling a complete simultaneous description of $\bar{q}q$ and $qq$ channels. As a result, the effective Lagrangian comprising $\bar{q}q$ interactions 
is supplemented with a corresponding $qq$ interacting part as follows:
\begin{equation}
\begin{aligned}
\mathcal{L}_\text{eff} &= \bar{q}\left(i \gamma^{\mu} \partial_{\mu}-m + \mu \gamma_{0} \right) q\\
&+\alpha \mathcal{F}_{\bar{q} q}(\mathcal{L}_{\mathrm{int}}^{(4)})+(1-\alpha) \mathcal{F}_{q q} (\mathcal{L}_{\mathrm{int}}^{(4)}). \label{eq3}  
\end{aligned}
\end{equation}    \\         
In Eq.~(\ref{eq3}), $\mathcal{F}_{\bar{q} q}(\mathcal{L}_{\mathrm{int}}^{(4)})$
represents the total Fierz-transformed quark-antiquark interactions, and $\mathcal{F}_{qq}(\mathcal{L}_{\mathrm{int}}^{(4)})$ indicates all diquark interactions after employing the Fierz transformation to the two-current interaction term in Eq.~(\ref{eq2}).

It is important to note that since the Fierz transformation and the mean-field approximation are non-commutative, changing $\alpha$ modifies the ratios of the diquark coupling constant to the quark-antiquark coupling constant, and the chiral condensates and diquark condensates influence each other in the mean-field approximation. We will discuss this point in more detail in the effects of changing $\alpha$ in Sec.~\ref{:result}.

\subsection{The interactions we considered at finite densities}
\label{subA}
In Eq.~(\ref{eq3}), the Fierz-transformed interactions contain various channels. 
In this section, we discuss which channels contribute to a nonzero expectation value at finite densities.
The explicit expressions of {$\bar{q} q$} channels and {$q q$} in Eq.~(\ref{eq3}) are listed in Eq.~(\ref{eq4}) and Eq.~(\ref{eq5}), respectively, 
\begin{widetext}
\begin{equation}
\begin{aligned}
\mathcal{F}_{\bar{q}q}(\mathcal{L}_{\mathrm{int}}^{(4)})=
&+
\frac{2\left(N_{c}^{2}-1\right)}{N_{c}^{2} N_{f}} g\left[(\bar{q} q)^{2}+\left(\bar{q} i \gamma_{5} q\right)^{2} \right.
\left.-\frac{1}{2}\left(\bar{q} \gamma^{\mu} q\right)^{2}-\frac{1}{2}\left(\bar{q} \gamma^{\mu} \gamma_{5} q\right)^{2}\right] 
+\frac{\left(N_{c}^{2}-1\right)}{N_{c}^{2}} g\left[\left(\bar{q} \tau_{n} q\right)^{2}+\left(\bar{q} i \gamma_{5} \tau_{n} q\right)^{2}\right.\\
&-\left.\frac{1}{2}\left(\bar{q} \gamma^{\mu} \tau_{n} q\right)^{2}-\frac{1}{2}\left(\bar{q} \gamma^{\mu} \gamma_{5} \tau_{n} q\right)^{2}\right] 
-\quad \frac{1}{N_{c} N_{f}} g\left[\left(\bar{q} \lambda_{a} q\right)^{2}+\left(\bar{q} i \gamma_{5} \lambda_{a} q\right)^{2} \quad \right.
\left.-\frac{1}{2}\left(\bar{q} \gamma^{\mu} \lambda_{a} q\right)^{2}-\frac{1}{2}\left(\bar{q} \gamma^{\mu} \gamma_{5} \lambda_{a} q\right)^{2}\right] \\
&- \frac{1}{2 N_{c}} g\left[\left(\bar{q} \lambda_{a} \tau_{n} q\right)^{2}+\left(\bar{q} i \gamma_{5} \lambda_{a} \tau_{n} q\right)^{2} \right. \left.-\frac{1}{2}\left(\bar{q} \gamma^{\mu} \lambda_{a} \tau_{n} q\right)^{2}-\frac{1}{2}\left(\bar{q} \gamma^{\mu} \gamma_{5} \lambda_{a} \tau_{n} q\right)^{2}\right],\\ \label{eq4}
\end{aligned}
\end{equation}
\end{widetext}
\begin{widetext}
\begin{equation}
\begin{aligned}
\mathcal{F}_{qq}(\mathcal{L}_{\mathrm{int}}^{(4)})=
&\frac{N_c+1}{2 N_c} g\left[\left(\bar{q} i \gamma_5 C \tau_A \lambda_{A^{\prime}} \bar{q}^T\right)\left(q^T C i \gamma_5 \tau_A \lambda_{A^{\prime}} q\right)+\left(\bar{q} C \tau_A \lambda_{A^{\prime}} \bar{q}^T\right)\left(q^T C \tau_A \lambda_{A^{\prime}} q\right)-\frac{1}{2}\left(\bar{q} \gamma^\mu \gamma_5 C \tau_A \lambda_{A^{\prime}} \bar{q}^T\right)\left(q^T C \gamma_\mu \gamma_5 \tau_A \lambda_{A^{\prime}} q\right)\right.\\
&\left.-\frac{1}{2}\left(\bar{q} \gamma^\mu C \tau_S \lambda_{A^{\prime}} \bar{q}^T\right)\left(q^T C \gamma_\mu \tau_S \lambda_{A^{\prime}} q\right)\right]-\frac{N_c-1}{2 N_c} g \left[ \left(\bar{q} i \gamma_5 C \tau_S \lambda_{S^{\prime}} \bar{q}^T\right)\left(q^T C i \gamma_5 \tau_S \lambda_{S^{\prime}} q\right)+\left(\bar{q} C \tau_S \lambda_{S^{\prime}} \bar{q}^T\right)\left(q^T C \tau_S \lambda_{S^{\prime}} q\right)\right.\\
&\left.-\frac{1}{2}\left(\bar{q} \gamma^\mu \gamma_5 C \tau_S \lambda_{S^{\prime}} \bar{q}^T\right)\left(q^T C \gamma_\mu \gamma_5 \tau_S \lambda_{S^{\prime}} q\right)-\frac{1}{2}\left(\bar{q} \gamma^\mu C \tau_S \lambda_{S^{\prime}} \bar{q}^T\right)\left(q^T C \gamma_\mu \tau_S \lambda_{S^{\prime}} q\right)\right]. \label{eq5}
\end{aligned}
\end{equation}
\end{widetext}
In Eq.~(\ref{eq4}), the notation $\tau_n$ and $\lambda_a$ corresponds to operators in the SU(2) flavor space and SU(3) color space, respectively. The expressions provided previously feature an implied summation over $a=$ $1, \ldots, N_{c}^{2}-1$ and $n=1, \ldots, N_{f}^{2}-1$. The interaction representation outlined in Eq.~(\ref{eq4}) holds the chiral symmetry of QCD at the tree level, facilitating the examination of the phase transitions of chiral symmetry breaking and restoration.
In Eq.~(\ref{eq5}), the superscript $T$ denotes transposition of the matrix of charge conjugation, which is represented as $C=\mathrm{i} \gamma^{2} \gamma^{0}$. Moreover, $\tau_{A,S}$ and $\lambda_{A,S}$ denote the antisymmetric or symmetric generators of the $\rm{SU(N_{f})}$ and $\rm{SU(N_{c})}$ groups, respectively, acting in the flavor space and color space. Per the restrictions imposed by the Pauli principle, the diquark condensates must be antisymmetric, and the diquark interactions that satisfy all the symmetry requirements are listed in Table.~\ref{table:1}.

\begin{table}
\centering
\caption{Dirac operators and generators of $\rm{U(2)}$ and $\rm{U(3)}$, and their symmetries under transposition. $\tau_{i} (i=1$-$3)$ denote Pauli matrices, and $\lambda_{i} (i=1$-$8)$ denote Gell-Mann matrices.}
         \vskip+2mm
\renewcommand\arraystretch{1.5}
\begin{ruledtabular}
\begin{tabular*}{\hsize}{@{}@{\extracolsep{\fill}}lcc@{}}
& Antisymmetric & Symmetric \\
\hline Dirac  & $C\gamma_5(S),C(P),C\gamma^{\mu}\gamma_5(V)$ & 
$C\gamma^{\mu}(A),C\sigma^{\mu\nu}(T)$\\
\hline $\rm{U(2)}$ & $\tau_2$(singlet) & $\mathbbm{1},\tau_1,\tau_3$(triplet) \\
\hline $\rm{U(3)}$ & $\lambda_2,\lambda_5,\lambda_7$(antitriplet) & 
$\mathbbm{1},\lambda_1,\lambda_3,\lambda_4,\lambda_6,\lambda_8$(sextet)
\end{tabular*}
\end{ruledtabular}
    \vspace{-0.4cm}
\label{table:1}
\end{table}
 
For the discussion of the ground state of QCD matter at finite baryon densities, several observations can be made:
\begin{itemize}

\item
As the chiral $\mathrm{SU(2)}$ symmetry is broken by the presence of bare quark mass in the QCD vacuum, it is natural to consider the chiral condensate $\langle \bar{q} q\rangle$ in the ground state of QCD matter. This condensate is directly related to the scalar color-singlet channel 
{($\bar{q}q)^2$} 
in Eq.~(\ref{eq4}) under mean-field approximation.

\item
In Eq.~(\ref{eq5}), a strong scalar attraction is observed between u and d quarks with antiparallel spins ($J^{P}=0^{+}$) in the color antitriplet channel. This interaction is described by the term $(\bar{q} i \gamma_5 C \tau_A \lambda_{A^{\prime}} \bar{q}^T)(q^T C i \gamma_5 \tau_A \lambda_{A^{\prime}} q)$. Following the ideas of BCS theory~\cite{1957PhRv..106..162B,1957PhRv..108.1175B}, it is reasonable to expect that a color superconductor's ground state would possess a nonzero expectation value in the attractive channels, such as the most notable diquark condensate $\delta_1=\langle q^{\mathrm{T}} C \gamma_{5} \tau_{2} \lambda_{A^{\prime}} q\rangle$.

\item
As one approaches finite densities, the vector interaction in Eq.~(\ref{eq4}) becomes consequential and leads to Lorentz noninvariant expectation values such as the density itself $\rho=\left\langle\bar{q} \gamma^{0}q\right\rangle$~\cite{2001PhRvD..65a4018B,2005PhRvD..72c4004R,2005PhR...407..205B}. When interactions conform to the Pauli principle, the two-flavor color superconductor's ground state should consider the diquark condensate $\delta_{2}=\left\langle q^{T} C \gamma^{0} \gamma_{5} \tau_{2} \lambda_{A^{\prime}} q\right\rangle$. This condensate originates from the vector diquark interaction $(\bar{q} \gamma^{\mu} \gamma_{5} C \tau_{A} \lambda_{A^{\prime}} \bar{q}^{T})(q^{T}C \gamma_{\mu} \gamma_{5} \tau_{A} \lambda_{A^{\prime}} q)$.

\item
The diquark condensates present in a color superconductor break the color gauge symmetry spontaneously via the Anderson-Higgs mechanism, resulting in the breaking of the color gauge group into an $\rm{SU(2)_c}$ subgroup. The conventional choice of the condensate pointing in the ``blue" direction indicates that only two colors (``red" and ``green") participate in the diquark condensate, implying the absence of color $\rm{SU(3)}$ invariance in such states. However, it is likely that the quark condensate receives different contributions from red (or green) and blue quarks, denoted by $\phi_{r}$ and $\phi_{b}$, respectively, leading to a non-vanishing expectation value $\phi_{8}=\left\langle\bar{q} \lambda_{8} q\right\rangle=\frac{2}{\sqrt{3}}\left(\phi_{r}-\phi_{b}\right)$ associated with the scalar color-octet interaction 
{$(\bar{q}\lambda_8 q)^2$}.
It's important to note that $\delta_{1}$ and $\delta_{2}$ leave a color $\rm{SU(2)}$ subgroup invariant, and therefore, all green quantities are identical to the red ones. Similarly, density differences are expected between red (or green) and blue quarks, where aside from the total number density $\rho =2\rho_{r} +\rho_{b}$, there could be a non-vanishing expectation value $\rho_{8}=\left\langle\bar{q} \gamma^{0} \lambda_{8} q\right\rangle=\frac{2}{\sqrt{3}}\left(\rho_{r}-\rho_{b}\right)$ associated with the vector color-octet interaction 
{$(\bar{q} \gamma^0 \lambda_8 q)^2$}.

\end{itemize}

At finite densities, only specific components of the interactions in Eq.~(\ref{eq3}) lead to non-vanishing expectation values under mean-field approximation. The relevant effective Lagrangian used to describe such interactions is given by:
\begin{equation}
\begin{aligned}
\mathcal{L}_{\mathrm{eff}} &=\frac{1}{2}\left[\bar{q}\left(i \gamma^\mu \partial_\mu-m+\mu \gamma^0\right) q+\bar{q}_c\left(-i \gamma^\mu \overleftarrow{\partial_\mu}-m-\mu \gamma^0\right) q_c\right] \\
&+\alpha\frac{N_c^{2}-1}{N_c^{2}}  g(\bar{q} q)^2-\alpha \frac{1}{2N_c} g \sum_{a=1}^8\left(\bar{q} \lambda_a q\right)^2 \\
&-\alpha \frac{N_c^{2}-1}{2N_c^2} g\left(\bar{q} \gamma^0 q\right)^2+\alpha \frac{1}{4N_c} g \sum_{a=1}^8\left(\bar{q} \gamma^0 \lambda_a q\right)^2 \\
&+(1-\alpha)\frac{N_{c}+1}{2N_{c}}g \left(\bar{q} i \gamma_5 \tau_A \lambda_{A^{\prime}} q_c\right)\left(\bar{q}_c i \gamma_5 \tau_A \lambda_{A^{\prime}} q\right) \\
&-(1-\alpha)\frac{N_{c}+1}{4N_{c}}g\left(\bar{q} \gamma^0 \gamma_5 \tau_A \lambda_{A^{\prime}} q_c\right)\left(\bar{q}_c \gamma_0 \gamma_5 \tau_A \lambda_{A^{\prime}} q\right)\ ,\label{:effLagrangain}
\end{aligned}
\end{equation} 
where the antisymmetric Pauli matrix $\tau_{A}$ and the generators $\lambda_{A^{\prime}}$ are $\tau_{2}$ and $\lambda_{2,5,7}$, respectively. 
Here, we define the effective coupling constants as follows:
\begin{equation}
\begin{aligned}
g_{s}^{(0)}=\alpha\frac{N_c^{2}-1}{N_c^{2}}  g\ ,\  & g_{s}^{(8)}=-\alpha \frac{1}{2N_c} g \ ,\\
g_{v}^{(0)}=-\alpha \frac{N_c^{2}-1}{2N_c^2} g\ , \ &
g_{v}^{(8)}=\alpha \frac{1}{4N_c} g \ ,\\
h_{1}=(1-\alpha)\frac{N_{c}+1}{2N_{c}}g \ , \ &
h_{2}=-(1-\alpha)\frac{N_{c}+1}{4N_{c}}g \ .\label{eq9}
\end{aligned}
\end{equation}
As an example, a coupling ratio of $h_1/g_s^{(0)}$ between 0.5 and 3 corresponds to a variation of $\alpha$ from 0.6 to 0.2. The commonly-used case is $h_1/g_s^{(0)}=3/4$ (or equivalently $\alpha=0.5$), as mentioned before.

\subsection{Mean-field approximation and massive quark propagator}
\label{subB} 
In order to determine the thermodynamic properties of quark matter, we transform the interactions into a bilinear form of quark fields using mean-field approximation. This approach is solvable, leading to the following effective Lagrangian:
\begin{equation}
\begin{aligned}
&\mathcal{L}_\text{eff}=\\
&+ \frac{1}{2}\left[\bar{q}\left(i \gamma^{\mu} \partial_{\mu}-M+\tilde{\mu} \gamma^{0}\right) q+\bar{q}_{c}\left(-i \gamma^{\mu} \partial_{\mu}-M-\tilde{\mu} \gamma^{0}\right) q_{c}\right] \\
&+\frac{1}{2}\left[\bar{q}_{c}\left(-\Delta_{1}^{*}\right) \gamma_{5} \tau_{2} \lambda_{A^{\prime}} q+\bar{q} \Delta_{1} \gamma_{5} \tau_{2} \lambda_{A^{\prime}} q_{c}\right] \\
&+\frac{1}{2}\left[\bar{q}_{c}\Delta_{2}^{*} \gamma_{0} \gamma_{5} \tau_{2} \lambda_{A^{\prime}} q+\bar{q} \Delta_{2} \gamma_{0} \gamma_{5} \tau_{2} \lambda_{A^{\prime}} q_{c}\right] \\
&-g_{s}^{(0)}\langle\bar{q} q\rangle^{2}-g_{s}^{(8)}\langle\bar{q} \lambda_{8} q\rangle^{2}- g_{v}^{(0)}\left\langle\bar{q} \gamma^{0} q\right\rangle^{2} - g_{v}^{(8)}\left\langle\bar{q} \gamma^{0}\lambda_{8} q\right\rangle^{2}\\
&-h_{1}\left\langle\bar{q} i \gamma_{5} \tau_{2} \lambda_{A^{\prime}} q_{c}\right\rangle\left\langle\bar{q}_{c} i \gamma_{5} \tau_{2} \lambda_{A^{\prime}} q\right\rangle\\
&-h_{2}\left\langle\bar{q} \gamma_{0}\gamma_{5} \tau_{2} \lambda_{A^{\prime}} q_{c}\right\rangle\left\langle\bar{q}_{c} \gamma_{0} \gamma_{5} \tau_{2} \lambda_{A^{\prime}} q\right\rangle,\label{eq10}
\end{aligned}
\end{equation}
in which we introduce the effective quark masses,
\begin{equation}
\begin{aligned}
M=& M_{0}+M_{8}\lambda_{8} \ ,\\
M_{0}=& m-2 g_{s}^{(0)} \phi \ , \quad M_{8}=-2 g_{s}^{(8)} \phi_{8} \ ,\label{eq11}
\end{aligned}
\end{equation}
the effective quark chemical potentials,
\begin{equation}
\begin{aligned}
\tilde{\mu}=&\mu_0+ \mu_8\lambda_{8}\ ,\\
\mu_0=&\mu+2 g_v^{(0)} \rho, \quad \mu_8=2 g_v^{(8)} \rho_8 \ ,
\end{aligned}
\end{equation}
and the diquark gaps,
\begin{equation}
\begin{aligned}
\Delta_{1}=&-2h_{1}\delta_{1}= -2 h_{1}\left\langle\bar{q}_{c} \gamma_{5} \tau_{2} \lambda_{A^{\prime}} q\right\rangle \ ,\\
\Delta_{2}=& 2h_{2}\delta_{2}=2 h_{2}\left\langle\bar{q}_{c}\gamma_{0} \gamma_{5} \tau_{2} \lambda_{A^{\prime}} q\right\rangle \ ,\\
\Delta_{1}^{*}=& 2h_{1}\left\langle\bar{q} \gamma_{5} \tau_{2} \lambda_{A^{\prime}} q_{c}\right\rangle \ , \\
\Delta_{2}^{*}=& 2h_{2}\left\langle\bar{q}\gamma_{0} \gamma_{5} \tau_{2} \lambda_{A^{\prime}} q_{c}\right\rangle \ , \\\label{eq13}
\end{aligned}
\end{equation}
with $q_{c}(x)=C \bar{q}^{T}(x)$ and $\bar{q}_{c}(x)= q^{T}(x) C$.

Next, we adopt an approach inspired by the BCS theory~\cite{1957PhRv..106..162B,1957PhRv..108.1175B}. This approach involves formally doubling the number of degrees of freedom by treating $q$ and $q_{c}$ as independent variables. We introduce the bispinor field $\Psi$ in the following manner:
\begin{equation}
\Psi(x)=\frac{1}{\sqrt{2}}
\left(\begin{array}{c}
 q(x) \\
 q_{c}(x)
\end{array}\right)\ .\label{eq7}
\end{equation}

Subsequently, the effective Lagrangian can be expressed in the momentum space as follows:
\begin{equation}
\mathcal{L}_\text{eff}=\bar{\Psi} S^{-1} \Psi+V \ ,\label{eq14}
\end{equation}
where $V$ is the interaction potential,
\begin{equation}
\begin{aligned}
V=&-g_{s}^{(0)}\langle\bar{q} q\rangle^{2}-g_{s}^{(8)}\langle\bar{q} \lambda_{8} q\rangle^{2}- g_{v}^{(0)}\left\langle\bar{q} \gamma^{0} q\right\rangle^{2} - g_{v}^{(8)}\left\langle\bar{q} \gamma^{0}\lambda_{8} q\right\rangle^{2}\\
&-h_{1}\left\langle\bar{q} i \gamma_{5} \tau_{A} \lambda_{A^{\prime}} q_{c}\right\rangle\left\langle\bar{q}_{c} i \gamma_{5} \tau_{A} \lambda_{A^{\prime}} q\right\rangle\\
&-h_{2}\left\langle\bar{q} \gamma_{0}\gamma_{5} \tau_{A} \lambda_{A^{\prime}} q_{c}\right\rangle\left\langle\bar{q}_{c} \gamma_{0} \gamma_{5} \tau_{A} \lambda_{A^{\prime}} q\right\rangle \ ,\label{eq15}
\end{aligned}
\end{equation}
and the inverse propagator of the $q$ fields at four-momentum $p$ is:
\begin{widetext}
\begin{equation}
S^{-1}(p)=\left(\begin{array}{cc}
\gamma^{\mu}p_{\mu}-M_{0}-M_{8} \lambda_{8}+\mu_{0} \gamma^{0}+\mu_{8} \gamma^{0} \lambda_{8} & 
\Delta_{1} \gamma_{5} \tau_{2} \lambda_{2}+\Delta_{2} \gamma^{0} \gamma_{5} \tau_{2} \lambda_{2} \\
-\Delta_{1}^{*} \gamma_{5} \tau_{2} \lambda_{2}+\Delta_{2}^{*} \gamma^{0} \gamma_{5} \tau_{2} \lambda_{2} & 
\gamma^{\mu}p_{\mu}-M_{0}-M_{8} \lambda_{8}-\mu_{0} \gamma^{0}-\mu_{8} \gamma^{0} \lambda_{8}
\end{array}\right) \ .\label{:propagator}
\end{equation}
\end{widetext}

To facilitate the interpretation of the results and for the sake of convenience, it is beneficial to perform linear combinations on the properties of red and blue quarks. For instance, we can express the red and blue constituent quark masses as $M_{r}=M_{0}+(1 / \sqrt{3}) M_{8}$ and $M_{b}=M_{0}-(2 / \sqrt{3}) M_{8}$, respectively. It is important to note that we have made an assumption of $\phi_{g}=\phi_{r}$ for the quark-antiquark condensate in order to preserve the unbroken $\rm{SU(2)_{c}}$ subgroup, as mentioned in Sec. \ref{subA}. Consequently, the following equation is obtained:
\begin{equation}
\begin{aligned}
M_{r}&=m-\frac{2}{3}\left(6 g_{s}^{(0)}+2 g_{s}^{(8)}\right) \phi_{r}-\frac{2}{3}\left(3 g_{s}^{(0)}-2 g_{s}^{(8)}\right) \phi_{b} \ , \\
M_{b}&=m-\frac{2}{3}\left(6 g_{s}^{(0)}-4 g_{s}^{(8)}\right) \phi_{r}-\frac{2}{3}\left(3 g_{s}^{(0)}+4 g_{s}^{(8)}\right) \phi_{b} \ ,\\
\tilde{\mu}_{r}&=\mu+\frac{2}{3}\left(6 g_{v}^{(0)}+2 g_{v}^{(8)}\right) \rho_{r}+\frac{2}{3}\left(3 g_{v}^{(0)}-2 g_{v}^{(8)}\right) \rho_{b} \ ,\\
\tilde{\mu}_{b}&=\mu+\frac{2}{3}\left(6 g_{v}^{(0)}-4 g_{v}^{(8)}\right) \rho_{r}+\frac{2}{3}\left(3 g_{v}^{(0)}+4 g_{v}^{(8)}\right) \rho_{b} \ .\label{: eff M mu}
\end{aligned}
\end{equation}
Equation (\ref{: eff M mu}) sheds light on the mechanism responsible for spontaneous chiral symmetry breaking, by which quarks obtain a dynamic mass, in addition to the minor impact by the bare quark mass. The vector interactions contribute to the expectation value of the quark number density, causing the physical chemical potential $\mu$ to shift to $\tilde{\mu}$.

\section{Thermodynamic Properties}
\label{:property}
In finite-temperature field theory~\cite{2011ftft.book.....K}, the linearization of $\mathcal{L}_\text{eff}$ in the vicinity of the expectation values and the application of Matsubara formalism yield the thermodynamic potential per volume of quark matter as follows:
\begin{equation}
\Omega(T, \mu)=-T \sum_{n} \int \frac{d^{3} p}{(2 \pi)^{3}} \frac{1}{2} \operatorname{Tr} \ln \left[\frac{1}{T} S^{-1}\left(i \omega_{n}, \boldsymbol{p}\right)\right]-V \ .\label{:Thermo}
\end{equation}
Here, $\omega_n$ represents the fermionic Matsubara frequencies, while $S^{-1}(p)$ refers to the inverse propagator of the $\Psi$ field at four-momentum $p$, which is given by the expression presented in Eq.~(\ref{:propagator}).
In the scenario of two flavors and three colors, the inverse propagator results in a $48 \times 48$ matrix. To calculate the trace in the equation for the thermodynamic potential per volume $\Omega(T, \mu)$, we must work within this 48-dimensional space. By performing the Matsubara sum, we obtain:
\begin{equation}
\begin{aligned}
\Omega(T, \mu)=&-4 \int \frac{d^{3} p}{(2 \pi)^{3}}\left\{2 \left(\frac{E_{+}+E_{-}}{2}+T \ln \left(1+e^{-E_{+} / T}\right)\right.\right.\\
&\left.+T \ln \left(1+e^{-E_{-} / T}\right)\right)+\left(\epsilon_{b}+T \ln \left(1+e^{-\epsilon_{+} / T}\right)\right.\\
&\left.\left.+T \ln \left(1+e^{-\epsilon_{-} / T}\right)\right)\right\}+V \ ,\label{:Thermo2}
\end{aligned}
\end{equation}
where physically irrelevant constant terms have been suppressed. 
In Eq.~(\ref{:Thermo2}), the 4 in front of the integral represents the spin and flavor degeneracy, while the 2 in the first line of the integrand takes into account the two paired colors. The second term in the large parentheses pertains to the blue quarks that do not partake in a diquark condensate, and their dispersion laws incorporated in this expression are the conventional ones:
\begin{equation}
\epsilon_{\pm}=\epsilon_{b} \pm \tilde{\mu}_{b}=\sqrt{\vec{p}^{2} +M_{b}^{2}} \pm \tilde{\mu}_{b} \ .\label{eq20}
\end{equation}
And the dispersion laws of red and green quarks are:
\begin{equation}
E_{\pm}=\sqrt{\vec{p}^{2}+M_{r}^{2}+\tilde{\mu}_{r}^{2}+\left|\Delta_{1}\right|^{2}+\left|\Delta_{2}\right|^{2} \pm 2 s} \ ,\label{:Dispersion}
\end{equation}
with
\begin{equation}
s=\sqrt{\left(\tilde{\mu}_{r}^{2}+\left|\Delta_{2}\right|^{2}\right) \vec{p}^{2}+t^{2}} \ , \quad t=M_{r} \tilde{\mu}_{r}-\operatorname{Re}\left(\Delta_{1} \Delta_{2}^{*}\right) \ .\label{eq22}
\end{equation}

In the physical scenario we are interested in, which is characterized by finite density and zero temperature, $\beta=1/T$ tends to infinity, resulting in
\begin{equation}
\ln \left(1+e^{-\beta x}\right) \to\,-\beta x \Theta(-x) \ ,\label{eq29}
\end{equation}
and the Fermi-Dirac distribution transforms into a step function, whereby:
\begin{equation}
n_{\mathrm{F}}(x)=\frac{1}{1+e^{\beta x}} \to\, \Theta(-x) \ . \label{eq30}
\end{equation} 
By utilizing these relations, we can simplify the grand thermodynamic potential mentioned in Eq.~(\ref{:Thermo2}). This simplification results in the following equation:
\begin{equation}
\begin{aligned}
\Omega(T, \mu) &=-4 \int \frac{d^{3} p}{(2 \pi)^{3}}\left\{\left(E_{+}+E_{-}\right)\right.\\
&\left.+\left[\epsilon_{b}-\epsilon_{+} \Theta\left(-\epsilon_{+}\right)-\epsilon_{-} \Theta\left(-\epsilon_{-}\right)\right]\right\}+V \\
&=-4 \int \frac{d^{3} p}{(2 \pi)^{3}}\left(E_{+}+E_{-}+\epsilon_{b}\right)\\
&+4 \int \frac{d^{3} p}{(2 \pi)^{3}} \epsilon_{-} \Theta\left(-\epsilon_{-}\right)+V \\
&=\Omega_{\mathrm{vacuum}}+\Omega_{b}+V \ ,\label{:final Z Thermo}
\end{aligned}
\end{equation}
with $\epsilon_{\pm}=\epsilon_{b} \pm \tilde{\mu}_{b}=\sqrt{p^{2}+M_{b}^{2}} \pm \tilde{\mu}_{b}$. 
Here, the $\Omega_{\mathrm{vacuum}}$ is the vacuum energy that needs to be regularized. 

So far, we have successfully derived the thermodynamic potential with finite chemical potential and zero temperature constraints. In order to ensure thermodynamic consistency, the condensates must be obtained through appropriate differentiation of the thermodynamic potential. The self-consistent solutions are those that correspond to the stationary points of the potential, which are defined by:
\begin{equation}
\frac{\delta \Omega}{\delta M_{0}}=\frac{\delta \Omega}{\delta M_{8}}=\frac{\delta \Omega}{\delta \mu_{0}}=\frac{\delta \Omega}{\delta \mu_{8}}=\frac{\delta \Omega}{\delta \Delta_{1}}=\frac{\delta \Omega}{\delta \Delta_{2}}=0 \ .\label{:gap eq}
\end{equation}
In cases where there are multiple stationary points, the stable solution is determined by the minimum value of $\Omega(T, \mu)$. Therefore, using Eq.~(\ref{:gap eq}), we can obtain the necessary expressions for the various expectation values as follows:
\begin{equation}
\begin{aligned}
\phi_{r}&=-4 \int \frac{d^{3} p}{(2 \pi)^{3}} \frac{1}{2 s}\left\{ \frac{1}{E_{+}}\left[M_{r} s+\tilde{\mu}_{r} t\right]+ \frac{1}{E_{-}}\left[M_{r} s-\tilde{\mu}_{r} t\right]\right\} \ ,\\
\phi_{b}&=-4 \int \frac{d^{3} p}{(2 \pi)^{3}} \frac{M_{b}}{\epsilon_{b}}\left[1-n\left(\epsilon_{-}\right)\right] \ ,\\
\rho_{r}&=4 \int \frac{d^{3} p}{(2 \pi)^{3}} \frac{1}{2 s}\left\{ \frac{1}{E_{+}}\left[\tilde{\mu}_{r}\left(s+\vec{p^2}\right)+M_{r} t\right]\right. \\
&\left.+ \frac{1}{E_{-}}\left[\tilde{\mu}_{r}\left(s-\vec{p^2}\right)-M_{r} t\right]\right\} \ ,\\
\rho_{b}&=4 \int \frac{d^{3} p}{(2 \pi)^{3}}n\left(\epsilon_{-}\right) \ ,\\
\delta_{1}&=-4 \int \frac{d^{3} p}{(2 \pi)^{3}} \frac{1}{s}\left\{ \frac{1}{E_{+}}\left[\Delta_{1} s-\Delta_{2} t\right]+ \frac{1}{E_{-}}\left[\Delta_{1} s+\Delta_{2} t\right]\right\},\\
\delta_{2}&=4 \int \frac{d^{3} p}{(2 \pi)^{3}} \frac{1}{s}\left\{ \frac{1}{E_{+}}\left[\Delta_{2}\left(s+\vec{p^2}\right)-\Delta_{1} t\right]\right. \\
&\left.+\frac{1}{E_{-}}\left[\Delta_{2}\left(s-\vec{p^2}\right)+\Delta_{1} t\right]\right\} \ .\label{:condensate}
\end{aligned}
\end{equation}

\section{Numerical results}
\label{:result}
In this section, we will investigate the phase structure at finite chemical potential through numerical calculations. Specifically, we will analyze the competition behavior between the chiral condensate and diquark condensate by varying the parameter $\alpha$. For a given $\alpha$ value, the other model parameters are adjusted to reproduce the QCD vacuum properties. Afterward, we set the bare quark mass $m$, the coupling constant $\alpha g$, and the cutoff $\Lambda$ to fit the pion mass, pion decay constant, and quark condensate. To do so, we adopt the parameter set from Ref.~\cite{1989NuPhA.504..668M}, where $m=5.5$~MeV, $g_s^{(0)}=5.074\times 10^{-6}\mathrm{MeV}^{-2}$, and the three-momentum cutoff $\Lambda=631$ MeV is used to regulate the ultraviolet divergences. It is worth mentioning that the presence of a small current quark mass $m$ induces partial restoration of chiral symmetry at large densities, leading to a small chiral condensate in the CSC phase. This nature has been referred to as the coexistence region~\cite{1999NuPhB.538..215B}.

\begin{figure}
\centering
\includegraphics[width=0.48\textwidth]{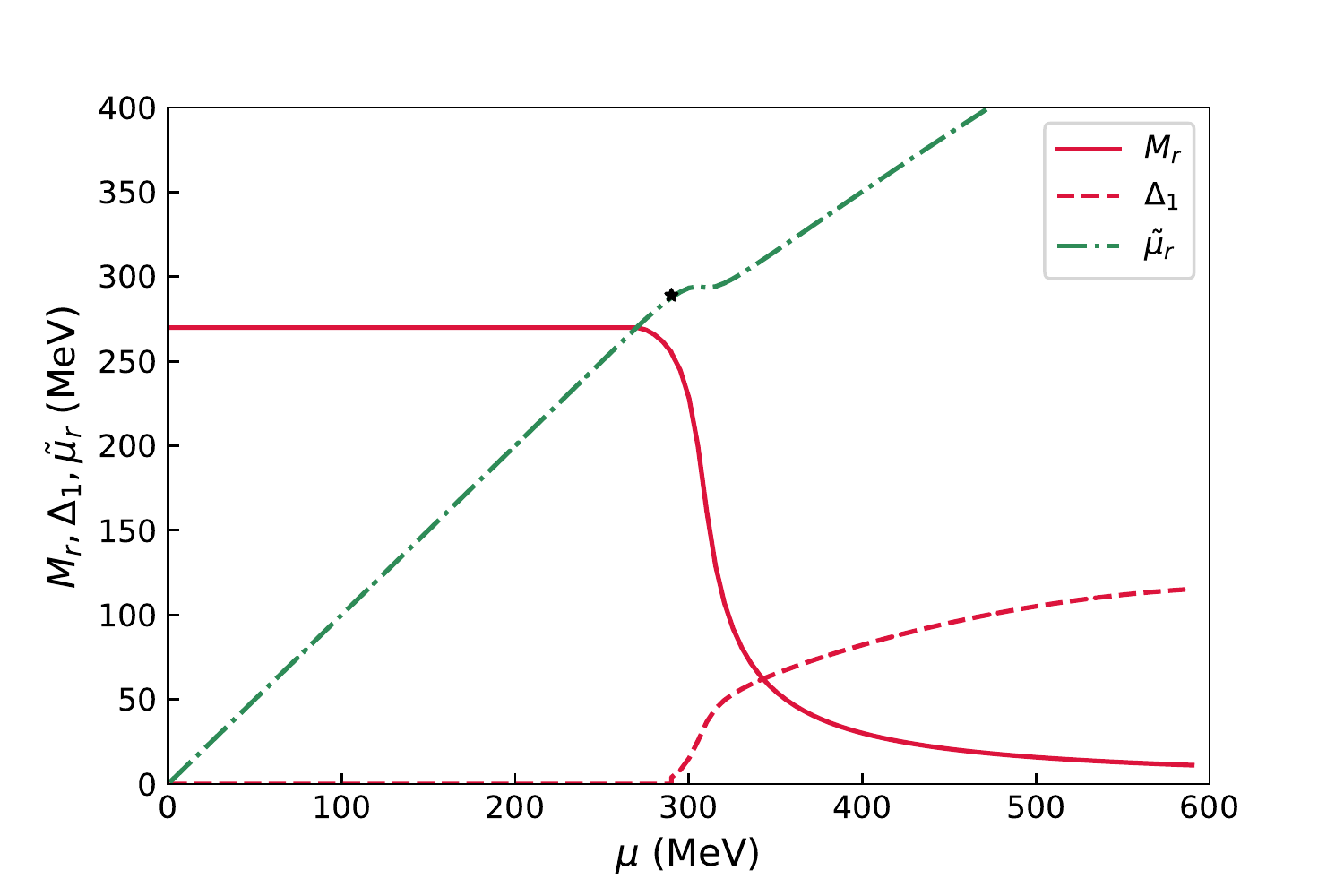}
\includegraphics[width=0.48\textwidth]{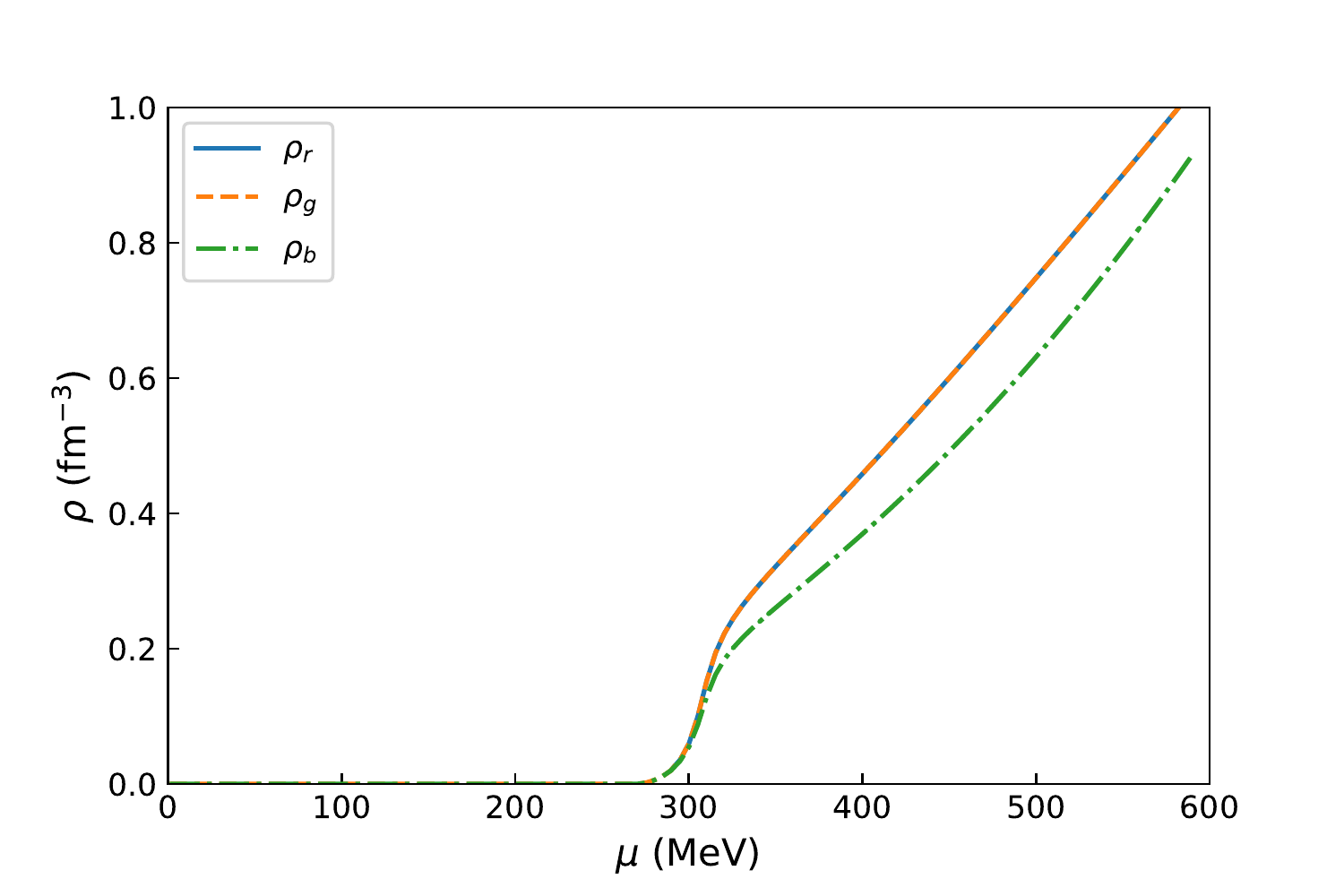}
\includegraphics[width=0.48\textwidth]{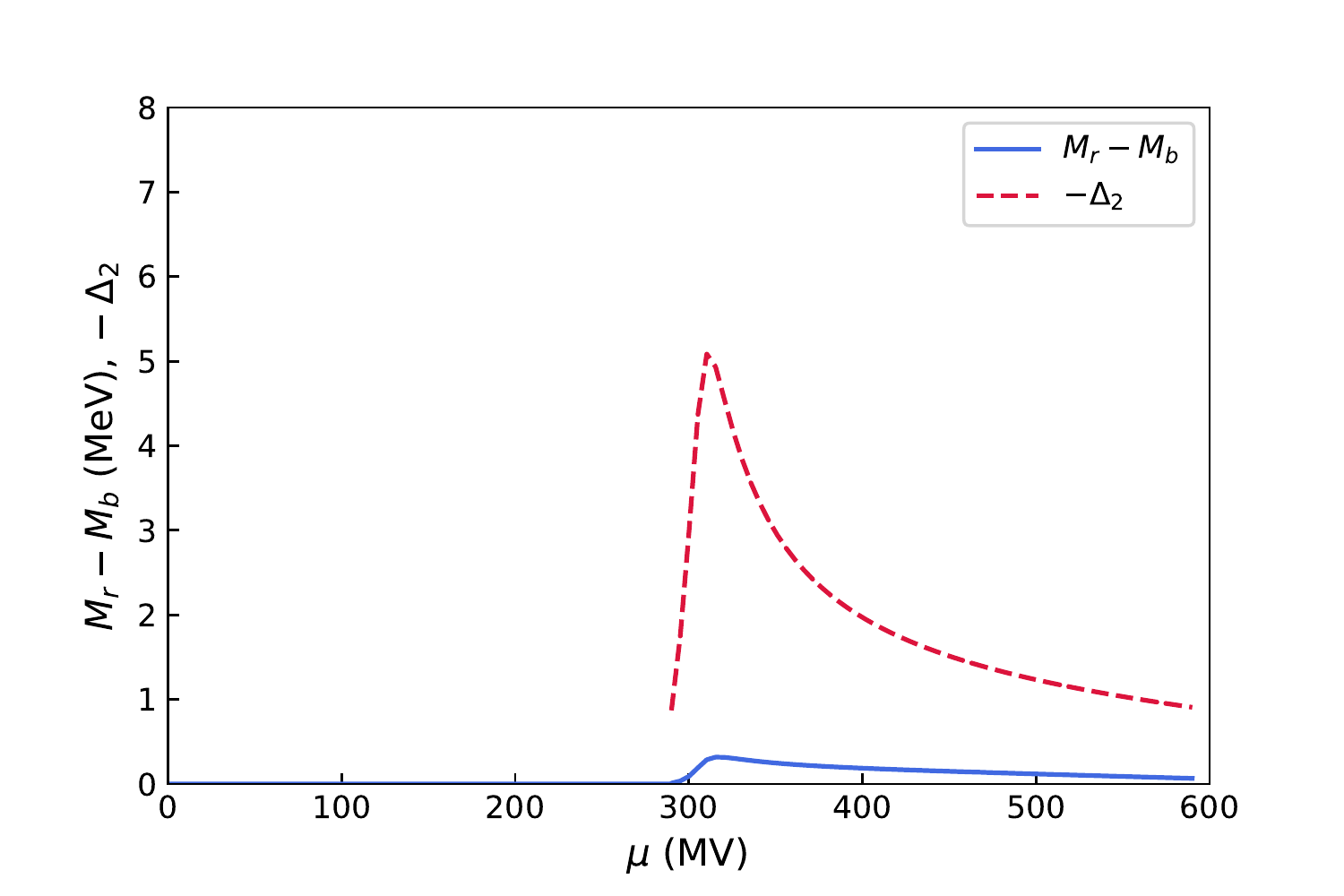}
     \vskip-2mm
\caption{Various quantities obtained as functions of the quark chemical potential for $\alpha=0.5$. Upper panel: Effective red quark mass $M_r$, effective red quark chemical potential $\tilde{\mu}_r$, and diquark gap $\Delta_1$ as functions of quark chemical potential.
The black star denotes the typical value of $\tilde{\mu}_r$ at which it first starts to deviate from the external chemical potential $\mu$. 
Middle panel: red, green, and blue quarks' number densities $\rho_{r,g,b}$ as functions of quark chemical potential.
Lower panel: $M_r-M_b$, $-\Delta_2$, as functions of quark chemical potential.  
}\label{fig:quantity 0.5}
    \vspace{-0.4cm}
\end{figure}

\begin{figure}
\centering
\includegraphics[width=0.45\textwidth]{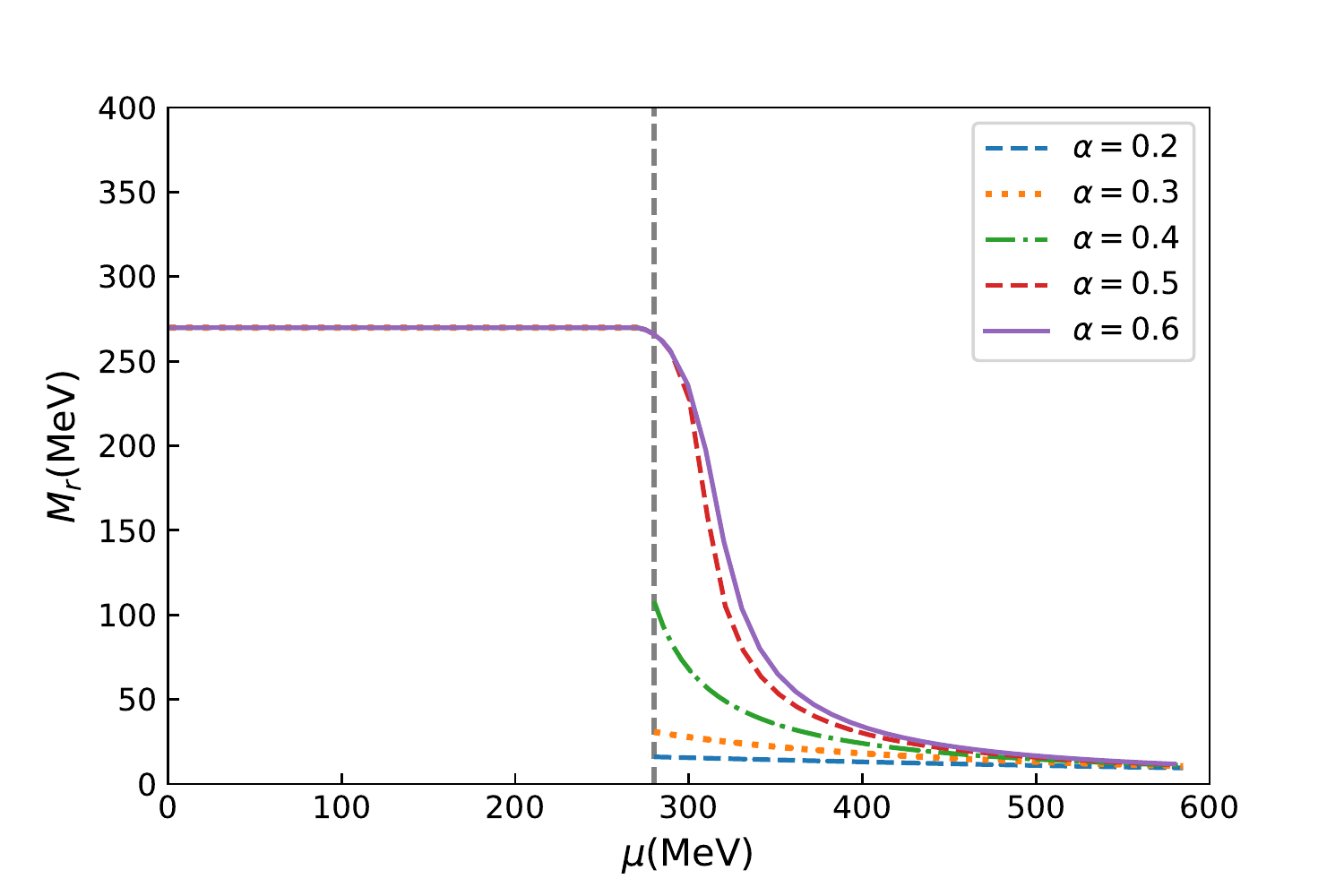}
    \vskip-2mm
\caption{Effective red quark mass $M_{r}$ as function of quark chemical potential for various selected values of $\alpha$. 
}\label{fig:M mu}
    \vspace{-0.4cm}
\end{figure}
 
We first present the results in the typical situation for $\alpha=0.5$, 
i.e., the relation between $h_1$ in the scalar diquark channel and $g_{s}^{(0)}$ in the scalar quark-antiquark channel is $3/4$, namely $h_1=3/4g_{s}^{(0)}$, 
as usually adopted by previous works~\cite{2001PhRvD..65a4018B,2005PhR...407..205B}. 
Fig.~\ref{fig:quantity 0.5} illustrates the behavior of various quantities, including the effective red quark mass $M_{r}$, the diquark gap $\Delta_1$, the effective red quark chemical potential $\tilde{\mu}_{r}$, the mass difference $M_r-M_b$, and the diquark gap $-\Delta_2$, as functions of the quark chemical potential $\mu$.
We observe that, for $\tilde{\mu}_{r} < M_{r}$, the effective quark mass remains constant at its vacuum value and the red (or green) quark number density remains zero. In this case, the effective quark chemical potentials $\tilde{\mu}_{r}$ are equal to the external chemical potential $\mu$, as predicted by Eq.~(\ref{: eff M mu}). As we increase $\mu$, chiral symmetry begins to restore, resulting in a decrease of the effective quark mass $M_{r}$, while the external chemical potential starts to exceed $\tilde{\mu}_{r}$. In the case of $\alpha=0.5$, the effective quark mass $M_{r}$ decreases smoothly, and the chiral phase transition is a crossover (see more related discussions of the chiral susceptibility below).
In a finite-density environment, the vector interactions contribute to the expectation value of the quark number density, modifying the quark chemical potential when $\mu$ surpasses the vacuum mass. As seen in Fig.~\ref{fig:quantity 0.5}, for $\tilde{\mu}_{r,b}>M_{r,b}$, the quark number density begins to increase smoothly from zero at a typical chemical potential.

Moreover, we find a scalar diquark gap $\Delta_1$ at around $\sim 100\mev$, which is consistent with previous studies~\cite{2001PhRvD..65a4018B,2005PhR...407..205B}. Similar to the BCS theory, an increase in the quark chemical potential leads to an increase in the density of states at the Fermi surface, resulting in a smooth growth of the gap parameter $\Delta_1$ when $\mu \gtrsim 290$~MeV. Since we consider the scalar color-cotet channels 
{$(\bar{q} \lambda_8 q)^2$}
in the Lagrangian [Eq.~(\ref{:effLagrangain})], the quark-antiquark condensates of red $\phi_r$ and blue $\phi_b$ quarks are expected to be different, leading to a small mass difference, as depicted in the lower panel of Fig.~\ref{fig:quantity 0.5}.
Meanwhile, the vector diquark gap parameter $\Delta_2$ becomes nonzero at $\mu \gtrsim 290\mev$. We will come back with a more detailed analysis of $\Delta_2$.

Next, we investigate the impact of varying the parameter $\alpha$, which corresponds to changing the ratio of $h_1/g_s^{(0)}$, on the competition between the quark-antiquark channels and diquark channels, and its possible impact on the phase transition. The results will be systematically discussed in the following section. For simplicity, we define the critical chemical potential $\mu_{\chi}$ for the chiral phase transition as the point at which the chiral condensate exhibits the maximum change, and we define the critical chemical potential $\mu_{\Delta_1}$ for the CSC phase transition as the point at which the diquark condensate starts to appear.

The effective red (or green) quark masses for different values of $\alpha$ are presented in Fig.~\ref{fig:M mu}. It can be observed that the quark masses experience a significant vacuum mass due to the spontaneous chiral symmetry breaking at $\mu=0$. However, up to a critical value of $\mu \sim 280\mev$, no further change in the quark masses is observed. Beyond this critical value, the chiral symmetry starts to restore.
By varying the parameter $\alpha$, the intensity of the scalar quark-antiquark fields is altered, which, in turn, modifies the characteristics of the chiral phase transition. A more detailed analysis reveals that for $\alpha=0.2,0.3,0.4$, a first-order phase transition occurs at $\mu_{\chi}\sim 280\mev$. However, as $\alpha$ is increased, the curves become smoother, and the chiral phase transition becomes a crossover. Nevertheless, no significant impact on the behavior of $M_{r}$ at finite quark chemical potential is observed at large $\alpha$, i.e., $\alpha \gtrsim 0.6$.

\begin{figure}
\centering
\includegraphics[width=0.49\textwidth]{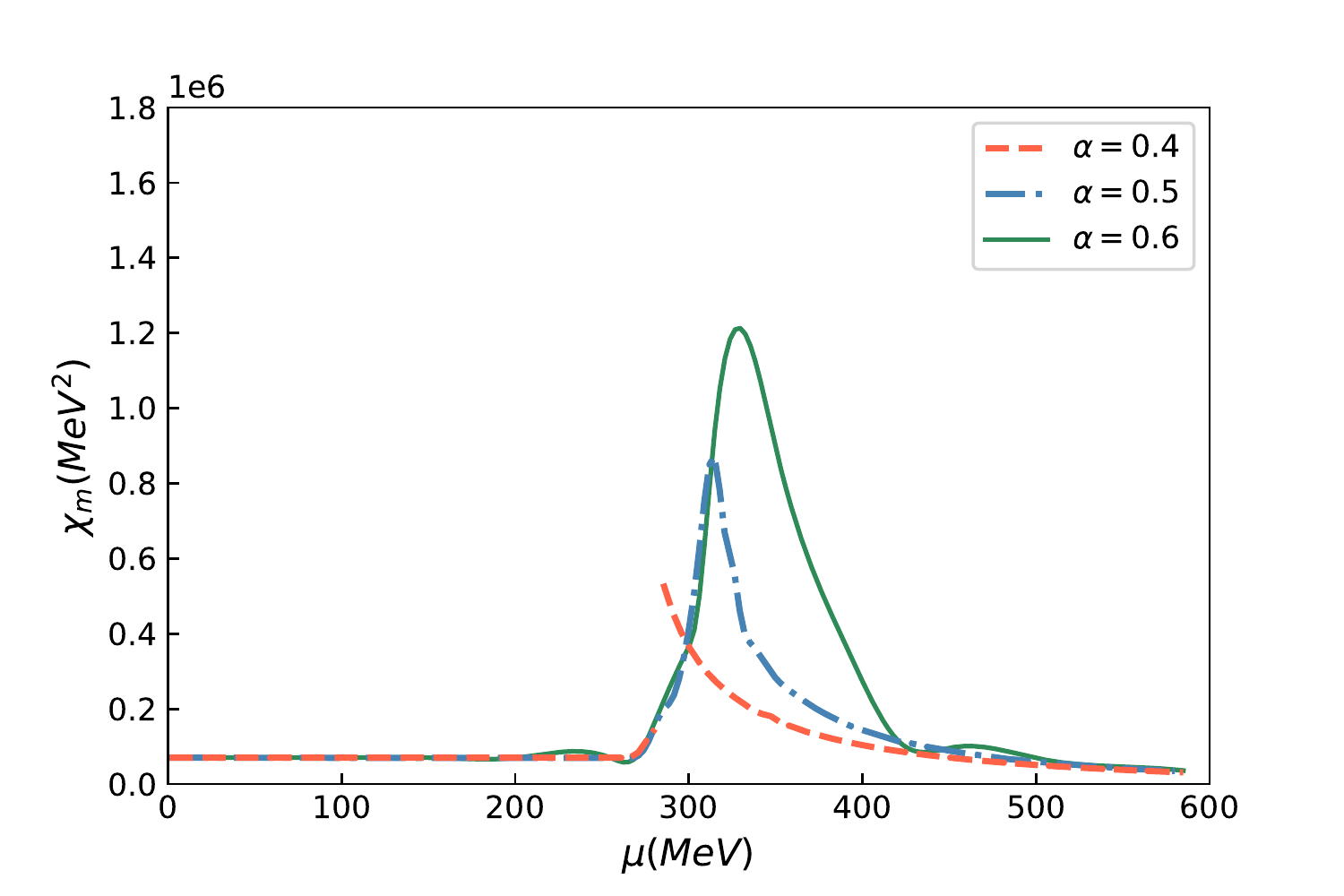}
    \vskip-2mm
\caption{Chiral susceptibility as function of quark chemical potential at zero temperature for $\alpha=0.4, 0.5, 0.6$, highlighting the transition from a first-order phase transition to a crossover.
}\label{fig:sus}
    \vspace{-0.4cm}
\end{figure}

To facilitate a clearer understanding of the chiral phase transition, we utilize the chiral susceptibility, defined as \begin{equation} \chi_r=\frac{\partial \sigma_r}{\partial m_r} \ .\label{:Chiral Sus} \end{equation} The resulting plot is depicted in Fig.~\ref{fig:sus}. The presence of a singular point on the susceptibility curve at $\alpha=0.4$, occurring at a critical chemical potential of $\mu_{\chi}=280$~\mev, indicates a first-order phase transition. On the other hand, for $\alpha$ exceeding a critical threshold of around $\alpha \gtrsim 0.5$, the curve appears smoother, pointing to a transition to the crossover phase.
 
\begin{figure}
\centering
\includegraphics[width=0.49\textwidth]{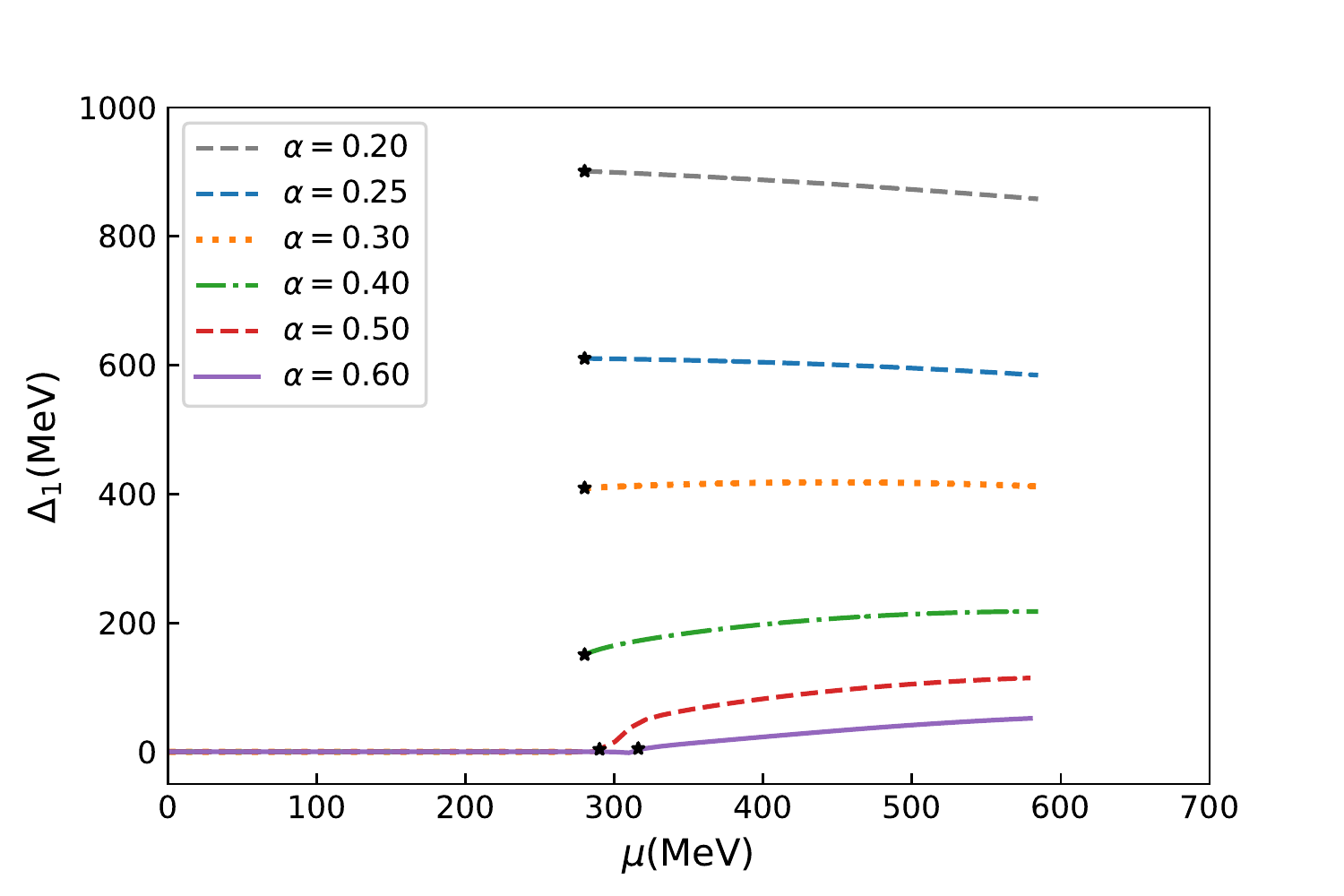}
    \vskip-2mm
\caption{Diquark gap $\Delta_1$ as function of quark chemical potential for various selected values of $\alpha$. 
}\label{fig:delta1 mu}
    \vspace{-0.4cm}
\end{figure}

\begin{figure}
\centering
\includegraphics[width=0.49\textwidth]{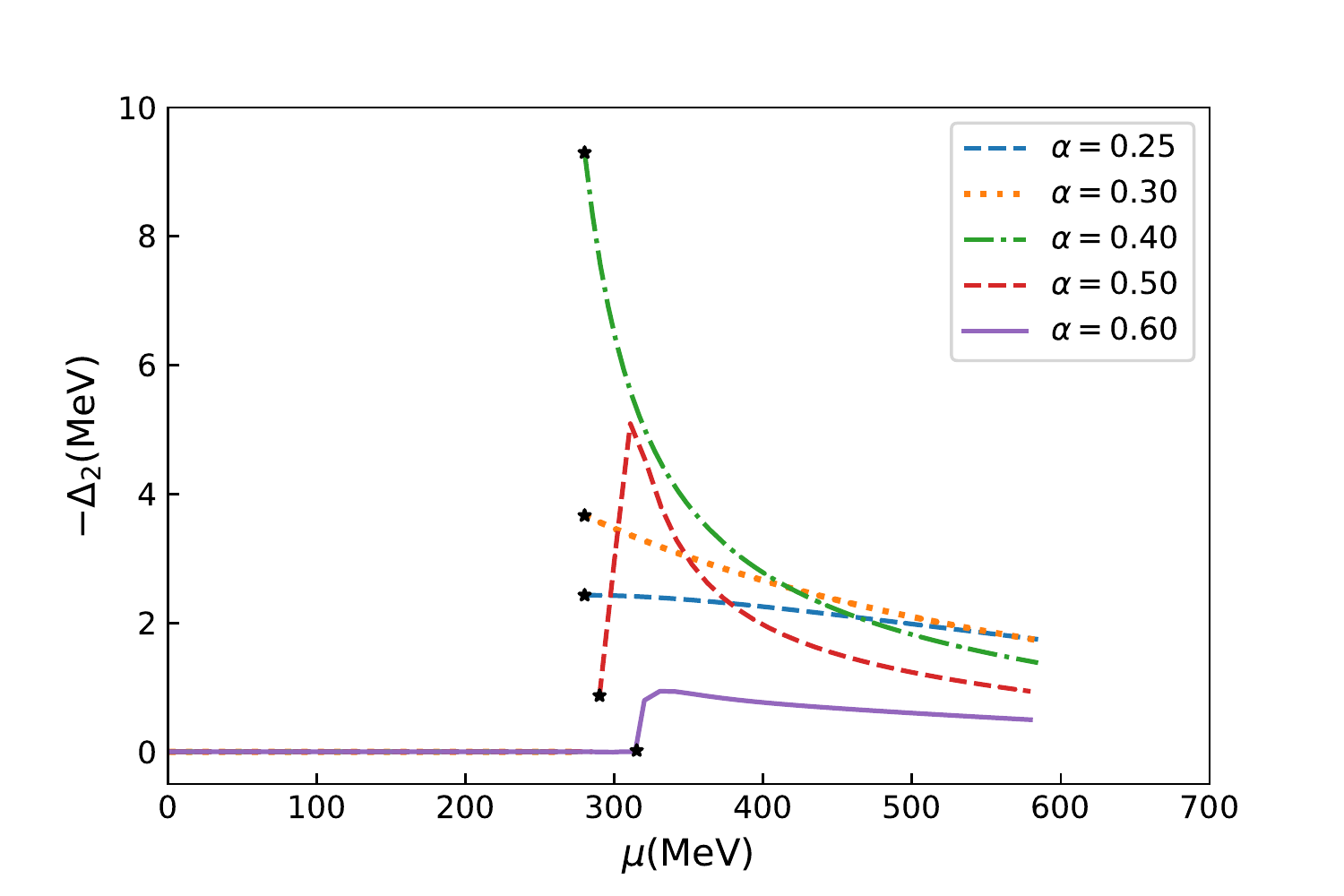}
    \vskip-2mm
\caption{Diquark gap $\Delta_2$ as function of quark chemical potential for various selected values of $\alpha$. 
}\label{fig:delta2 mu}
    \vspace{-0.4cm}
\end{figure}

One interesting finding among our results is the diquark gap $\Delta_1$, which is presented in Fig.~\ref{fig:delta1 mu}. 
 Specifically, at $\alpha=0.2$, the scalar diquark gap parameter $\Delta_1$ assumes a value exceeding the effective energy scale of the chosen parameters in NJL model, reaching nearly 850 MeV. However, for $\alpha \geq 0.25$, the corresponding gap parameter $\Delta_1$ stabilizes at a more reasonable value of approximately $610$~\mev. Consequently, we restrict our discussions to the range of parameter values starting from $\alpha=0.25$ (or equivalently $h_1/g_s^{(0)}=2.25$) onwards.
For $\alpha=0.3$ and $0.4$, we observe that the diquark condensate remains zero in the region where the constituent quark mass retains its value in the vacuum. As we increase the quark chemical potential, both the chiral and CSC phase transitions occur nearly simultaneously near the critical chemical potential $\mu_{\chi}=\mu_{\Delta_1}\sim 280$~\mev, with both transitions being of the first order.
When $\alpha=0.5$ and $0.6$, the chiral phase transition shifts to a crossover, and the scalar diquark condensates $\Delta_1$ demonstrate a continuous increase from zero.

The influence of the strength of the diquark coupling constant on the properties of the chiral and superconducting phase transitions can be explained using Eq.~(\ref{:effLagrangain}). 
Smaller values of $\alpha$ ($\alpha \lesssim 0.5$) correspond to stronger coupling constants in the diquark channels. Accordingly, the ratio of $h_1/g_s^{(0)}$ is large, causing the scalar diquark condensate $\Delta_1$ to escalate rapidly from zero to a non-zero value (e.g., $\sim 400$~MeV for $\alpha=0.3$, with $h_1/g_s^{(0)}=1.75$), as observed in Fig.~\ref{fig:delta1 mu}.
Previous NJL calculations with scalar diquark interaction found similarly large $\Delta_1$ diquark gap~\cite{2002PhRvD..65g6012H}.
In contrast, for $\alpha=0.5, 0.6$, the critical chemical potentials for chiral phase transition are $\mu_{\chi}\sim 315$~MeV and $\mu_{\chi} \sim 325$~MeV in Fig.~\ref{fig:M mu}, and the scalar diquark condensates $\Delta_1$ only appear around $\sim 290$~MeV and $\sim 316$~MeV, respectively, as shown in Fig.~\ref{fig:delta1 mu}. These findings suggest that greater magnitudes of diquark condensates (smaller values of $\alpha$) not only lead to the maximum gap $\Delta_1$ becoming greater, but also to the diquark condensates arising at a lower chemical potential. Additionally, the region of mixed broken phase $\mu_{\chi}-\mu_{\Delta_1}$ is wider for $\alpha=0.5$ than for $\alpha=0.6$. When $\mu < \mu_{\Delta_1}$, chiral symmetry is broken. In the region between $\mu_{\Delta_1}$ and $\mu_{\chi}$, both chiral and color symmetries are broken. When $\mu> \mu_{\chi}$, partial restoration of chiral symmetry occurs, and color superconductivity dominates.
Moreover, for $\alpha$ exceeding $0.6$, the attractive diquark interaction channel is weak, and the diquark condensates become too small to observe.

Fig. \ref{fig:delta2 mu} illustrates the behavior of the vector diquark condensate $\Delta_2$, which has a sign opposite to $\Delta_1$ and is more than one order of magnitude smaller. For $\alpha=0.3$ and $0.4$, the magnitude of $-\Delta_2$ decreases as chemical potential increases. The behavior of $\Delta_2$ as a function of chemical potential, shown in Fig. \ref{fig:delta2 mu}, is closely connected to the nature of the phase transition. Specifically, for $\alpha=0.3$ and $0.4$, the chiral phase transition is a first-order transition, as previously observed in Fig.~\ref{fig:M mu} and Fig.~\ref{fig:sus}; accordingly, $-\Delta_2$ jumps from zero to a non-zero value at the threshold chemical potential. For $\alpha=0.5$ and $0.6$, when the chiral phase transition becomes a crossover, non-zero values of $-\Delta_2$ also appear near $\sim 290$~MeV and $\sim 316$~MeV, respectively, in agreement with the corresponding results for $\Delta_1$.

Our results indicate that the parameter $\alpha$ has a significant impact on the characteristics of the phase transition, which arise from the interplay between the {$\bar{q}q$} and {$qq$} channels.
A higher value of $\alpha$ corresponds to a lower $h_1/g_s^{(0)}$ ratio, resulting in a weakened diquark coupling strength $h_1$ at the fermi surface. Consequently, the diquark gap $\Delta_1$ decreases, and both the chiral phase transition and color superconductivity phase transition become crossovers from first-order transitions. This phenomenon is reminiscent of the findings in the NJL model when studying the chiral phase transition without diquark condensates: a first-order phase transition transforms into a second-order transition or a smooth crossover when the coupling strength in the repulsive vector channel exceeds a certain value~\cite{1996NuPhA.611..393B}.

\section{Summary and conclusion}
\label{:summary}
The existence of the CSC phase at finite densities makes the QCD phase diagram more complex and fascinating. In this study, we employ the Fierz transformation to incorporate diquark interactions self-consistently within a modified NJL-type model. Through mean-field approximation, we analyze the chiral condensate and the diquark condensate on equal footing in the two-flavor case, while also discussing the order of the chiral phase transition by examining the chiral susceptibility.

We observe that the nature of the phase transition and the magnitude of the diquark gap depend on the relative strengths of {$\bar{q}q$} and {$qq$}
interacting channels, which are quantified by the parameter $\alpha$ assigned to each channel. Our investigation reveals that for $\alpha$ values exceeding $0.25$, unphysically large diquark gaps are obtained, pointing to a restriction on the maximum allowed value of $\alpha$. Within the range of $\alpha$ values ranging from 0.25 to 0.6, changes in the relative strength are observed, giving rise to a transformation of both the chiral and CSC phase transitions from first order to crossover. When $\alpha$ surpasses 0.6, the effective mass versus chemical potential relationship becomes indistinguishable, and the diquark gap becomes too small to be detectable.

We conducted a detailed analysis of the competition mechanism among different phases. In our model, the chiral condensate and diquark condensate compete for the same quark, with the larger of the two condensates suppressing the smaller one. Thus, smaller $\alpha$, or a larger magnitude of $h_1/g_s^{(0)}$, promotes the formation of CSC and increases the diquark gap $\Delta_1$. Our results indicate a physical magnitude of scalar diquark gap $\Delta_1$ of approximately $400 \mev$ or greater, while the vector diquark condensate $\Delta_2$ is found to be small. The presence of diquark condensates leads to spontaneous breaking of the color gauge symmetry, resulting in slight distinctions between the condensates of red and blue quarks involved in pairing. However, this effect is found to be negligible in our study.

To gain a better understanding of the CSC phase, future research within this framework must constrain the range of $\alpha$. This constraint is particularly important when contrasting theoretical results with available observations, such as those concerning CSC compact stars~\cite{2011RAA....11..482L,2017ApJ...844...41L,2018PhRvD..97h3015Z,2021ApJ...917L..22M,2021MNRAS.506.5916L}. For such purposes, neutral conditions must be imposed since the stellar matter will be in beta equilibrium. Additionally, despite the heavier mass of strange quarks compared to up and down quarks, their appearance in the core of compact stars remains a possible scenario. Comprehensive investigations in this area are already underway.


\medskip
\acknowledgments
We are thankful to Yong-Hui Xia, Jia-Ning Li, Bo-Lin Li, Wei-Jie Fu, Yan Yan, and the XMU neutron star group for helpful discussions. 
The work is supported by the National SKA Program of China (No.~2020SKA0120300) and the National Natural Science Foundation of China (Grant Nos.~12273028 and 11873040).

\end{document}